\documentclass[a4paper,fleqn,usenatbib]{mnras}

\usepackage[T1]{fontenc}
\usepackage{ae,aecompl}

\usepackage{graphicx}	
\usepackage{amsmath}	
\usepackage{amssymb}	
\usepackage{multirow}      
\usepackage{scrextend}    

\title[SPRAI - radiative feedback coupled with primordial chemistry]{SPRAI: Coupling of radiative feedback and primordial chemistry in moving mesh hydrodynamics}

\author[O. Jaura et al.]{
Ondrej Jaura,$^{1}$\thanks{E-mail: ondrej.jaura@uni-heidelberg.de (ITA ZAH)},
S. C. O. Glover$^{1}$,
R. S. Klessen$^{1,2}$,
J.-P. Paardekooper$^{1}$
\\
$^{1}$Zentrum f\"{u}r Astronomie der Universit\"{a}t Heidelberg, Institut f\"{u}r Theoretische Astrophysik, Albert-Ueberle-Str. 2, 69120 Heidelberg, Germany\\
$^{2}$Universit\"{a}t Heidelberg, Interdisziplin\"{a}res Zentrum f\"{u}r Wissenschaftliches Rechnen, INF 369, 69120 Heidelberg \\
}

\date{Accepted XXX. Received YYY; in original form ZZZ}

\pubyear{2015}

\begin{document}
\label{firstpage}
\pagerange{\pageref{firstpage}--\pageref{lastpage}}
\maketitle
\begin{abstract}
In this paper we introduce a new radiative transfer code SPRAI (Simplex Photon Radiation in the Arepo Implementation) based on the SimpleX radiation transfer method. This method, originally used only for post-processing, is now directly integrated into the Arepo code and takes advantage of its adaptive unstructured mesh. Radiated photons are transferred from the sources through the series of Voronoi gas cells within a specific solid angle. From the photon attenuation we derive corresponding photon fluxes and ionization rates and feed them to a primordial chemistry module. This gives us a self-consistent method for studying dynamical and chemical processes caused by ionizing sources in primordial gas. Since the computational cost of the SimpleX method does not scale directly with the number of sources, it is convenient for studying systems such as primordial star-forming halos that may form multiple ionizing sources. 
\end{abstract}




\section{Introduction}

Radiation feedback plays an important role in cosmic structure formation on small scales (e.g.\ protostellar disk fragmentation during Population III star formation; see for example \citealt{Stacy2016} or \citealt{Hosokawa2016}) as well as on large scales (e.g.\ re-ionization of the universe; see e.g.\ \citealt{Paardekooper2015}, \citealt{Pawlik2016}, or \citealt{Xu2016a}). Most modern hydrodynamical codes use either ray tracing, moment-based or Monte Carlo methods to calculate the ionizing radiation field in a simulated medium. However, since the computational cost of most methods scales linearly with the number of sources, simulations with many sources are computationally very expensive. Therefore, in many cases radiation is treated either in a post-processing step \citep{Paardekooper2015}, or it is coupled to hydrodynamics, but with just a small number of sources \citep{Hosokawa2016, Stacy2016}. Use of such approximations in simulations neglects local dynamical and ionization variations in the gas that can be properly resolved only for multi-frequency and multi-source ray tracing. 

There are many codes and implementations available that are able to follow the ionizing radiation from different astrophysical sources. Properties and performance of the most widely used codes are summarised in code comparison projects by \citet{Iliev2006, Iliev2009} and \citet{Bisbas2015}. One of the codes that is commonly used in computational astrophysics is the moving mesh hydrodynamical code Arepo \citep{Springel2010}. Because of its versatility, it has been used in several large scale cosmological simulation projects like Illustris \citep{Vogelsberger2014}, Auriga \citep{Grand2017} or IllustrisTNG \citep{Pillepich2017a}, as well as numerous small-scale simulations, ranging from the formation of the first stars \citep{Greif2011} to the formation of structures in present-day molecular clouds \citep{Smith2016}. Although various efforts have been made to add radiation transfer to Arepo 
\citep[e.g.][]{Petkova2011,Greif2014,Bauer2015}, none of the current implementations is suitable for computationally challenging multi-source simulations where radiation feedback plays an important part in structure formation. Since our future work in the field of first star formation will involve such environments, it is our priority to develop a radiative transfer method that will address this challenge. 

In this paper we introduce an alternative method for ray tracing, SimpleX, that we implement in Arepo. The SimpleX method was originally developed for use with static Voronoi grids and is described and analyzed in \citet{ritzerveld2006TraAdaRanLat}, \citet{Paardekooper2010} and \citet{Kruip2010}. Our implementation is the first time that the method has been used on-the-fly within a hydrodynamical simulation, and takes advantage of the moving mesh already present during the hydrodynamical calculations of the Arepo code. This makes the combination of the two concepts easier and more efficient.

A characteristic feature of the SimpleX method is that the computational cost does not scale with the number of sources, as is the case for conventional ray-tracing algorithms. Sources in SimpleX only inject new photons into corresponding grid cells. All photons on the grid are subsequently transferred from cell to cell in a sequence of radiation transfer steps. From the movement of photons one can than calculate the corresponding fluxes and ionisation rates in each cell. The collective transfer of photons during the radiation transfer makes this method especially useful in simulations with a high number of sources. 

Our paper is structured as follows. In Section 2, we describe details of the SimpleX method and its coupling to a primordial chemistry network. Section 3 shows how the code performs for some standard test problems (e.g.\ the growth of the ionization front in uniform density gas in the R-type or the D-type regime). In the final section, we summarise our results and give an overview of our future plans. 

\section{Description of the method}

SPRAI is based on the SimpleX radiative transfer method developed by \citet{ritzerveld2006TraAdaRanLat}, \citet{Paardekooper2010} and \citet{Kruip2010} and used in \citet{Paardekooper2015}. This method was previously used only for post-processing results from hydrodynamical simulations of cosmic structure formation. The original SimpleX code uses its own integrated chemistry model that includes ionization, recombination and cooling of hydrogen and helium in several frequency bins.

Our implementation of the radiation transfer (RT) of ionizing photons in Arepo \citep{Springel2010} is based on the updated version of the original SimpleX algorithm discussed in \citet{Paardekooper2010}. In our implementation, the photon transfer is now calculated directly on the Voronoi mesh created in Arepo, rather than on a mesh created during a post-processing step. From the photon transfer, we obtain photon fluxes in each cell, and from these we can derive the corresponding ionization and heating rates. These rates are subsequently used by a separate chemistry module to update the chemical and thermal state of the gas. In the following sections we will introduce physical and numerical aspects of the radiation transfer as well as the chemistry module.

\subsection{Photon transfer}

The original SimpleX algorithm was discussed in \citet{Paardekooper2010} and mathematical analysis of the method\footnote{The transfer method that we use is a modified version of the direction-conserving transport (DCT) method discussed in the paper of \citet{Kruip2010}} was studied by \citet{Kruip2010}. In this section we will focus on our implementation which is adapted to work with the Arepo code.

Arepo is a multipurpose and multi-physics hydrodynamical code developed by \citet{Springel2010} and it is well established in the astrophysical community. In comparison with conventional Eulerian codes with static or adaptive meshes, Arepo uses a Voronoi mesh that is constructed at each step from a set of mesh generating points that move with the gas flow. Once the Voronoi mesh is constructed, Arepo calculates fluxes and gradients between cells and updates the cell values accordingly. An advantage of this method is that such a Voronoi mesh follows the underlying gas structure and effectively reduces grid artefacts that can arise from conventional mesh codes. Besides that, Arepo uses adaptive time stepping that reduces computational time of the simulation.

\begin{figure}
  \includegraphics[width=\columnwidth,trim={0.8cm 0.6cm 0.5cm 0.8cm},clip]{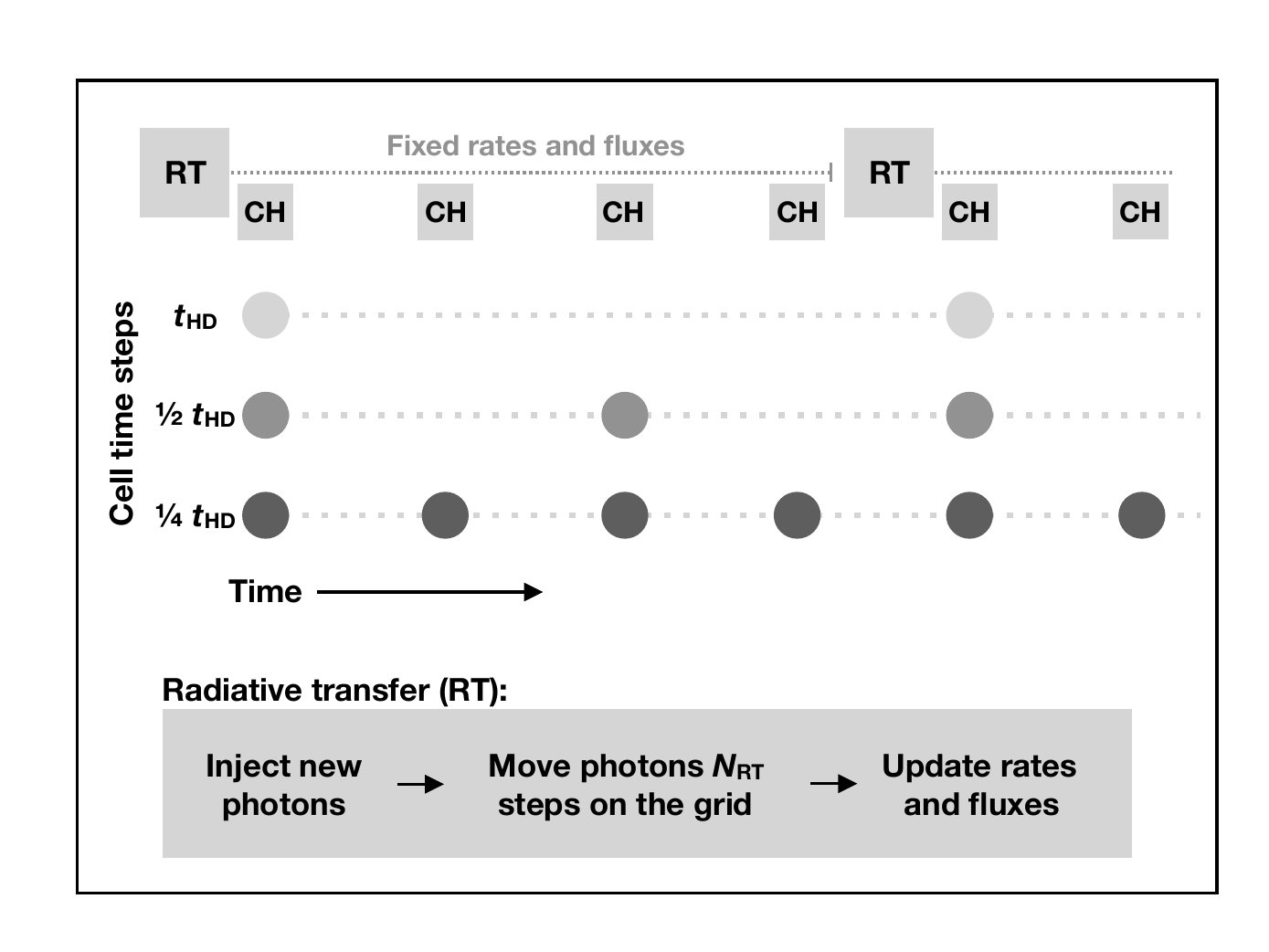}
    \caption{Schematic structure of the SPRAI implementation. Radiative transport (RT) is computed at the end of each full hydrodynamical step $t_{\mathrm{HD}}$, when all cells are synchronised and the complete Voronoi mesh is available. The calculated rates and fluxes are stored and used by the chemistry module (CH) during subsequent hydrodynamical steps.}
    \label{fig:code_scheme}
\end{figure}

In Figure \ref{fig:code_scheme} we show a schematic description of our implementation. Since construction and maintenance of the Voronoi mesh is already an integral part of the Arepo code, it can be easily used for SimpleX photon transport calculations. A complete mesh is constructed only on time steps when of all gas cells are synchronised. This means that the radiation transport in all cells can occur only at full hydrodynamical (HD) steps $t_\mathrm{HD}$. During HD sub-steps when only some cells are active we keep the ionization rates and radiation fluxes fixed\footnote{In principle one could also use a partially constructed mesh during HD sub-steps for radiative transport. However, this introduces several nontrivial challenges to the current algorithm that we will try to address in the future.}. 

At the beginning of each RT calculation we find all sources (ideal sources, stars, sink particles, ...) and calculate their immediate photon emission rates\footnote{In Section \ref{sec:TestsAndResults}, we use only ideal monochromatic sources with fixed ionization rates. However, the method can easily be extended to handle more realistic source spectra.}. For each source we find the nearest gas cell and assign to it the number of photons emitted by the source during $t_\mathrm{HD}$. Then we move photons on the Voronoi grid from cell to cell and calculate the ionization rates and radiation fluxes due to these photons as described in Section \ref{section:radiation_field}. One RT step is a movement of photons from one cell to a neighbouring cell. The number of RT steps $N_{\mathrm{RT}}$ depends on the ionization state of the gas through which the photons are passing. In general, the photons are propagated until they are completely attenuated in the gas.

It is important to note that in the present version of the method, we do not constrain the distance that the photons can propagate during $t_\mathrm{HD}$, i.e.\ we treat the speed of light as if it were infinite. For many applications, this is a reasonable approximation, but it may potentially cause problems in large cosmological simulations with the box sizes $L \gg c \; t_\mathrm{HD}$, where c is the speed of light.

\begin{figure}
  \includegraphics[width=\columnwidth,trim={1.85cm 1.3cm 1.5cm 1.3cm},clip]{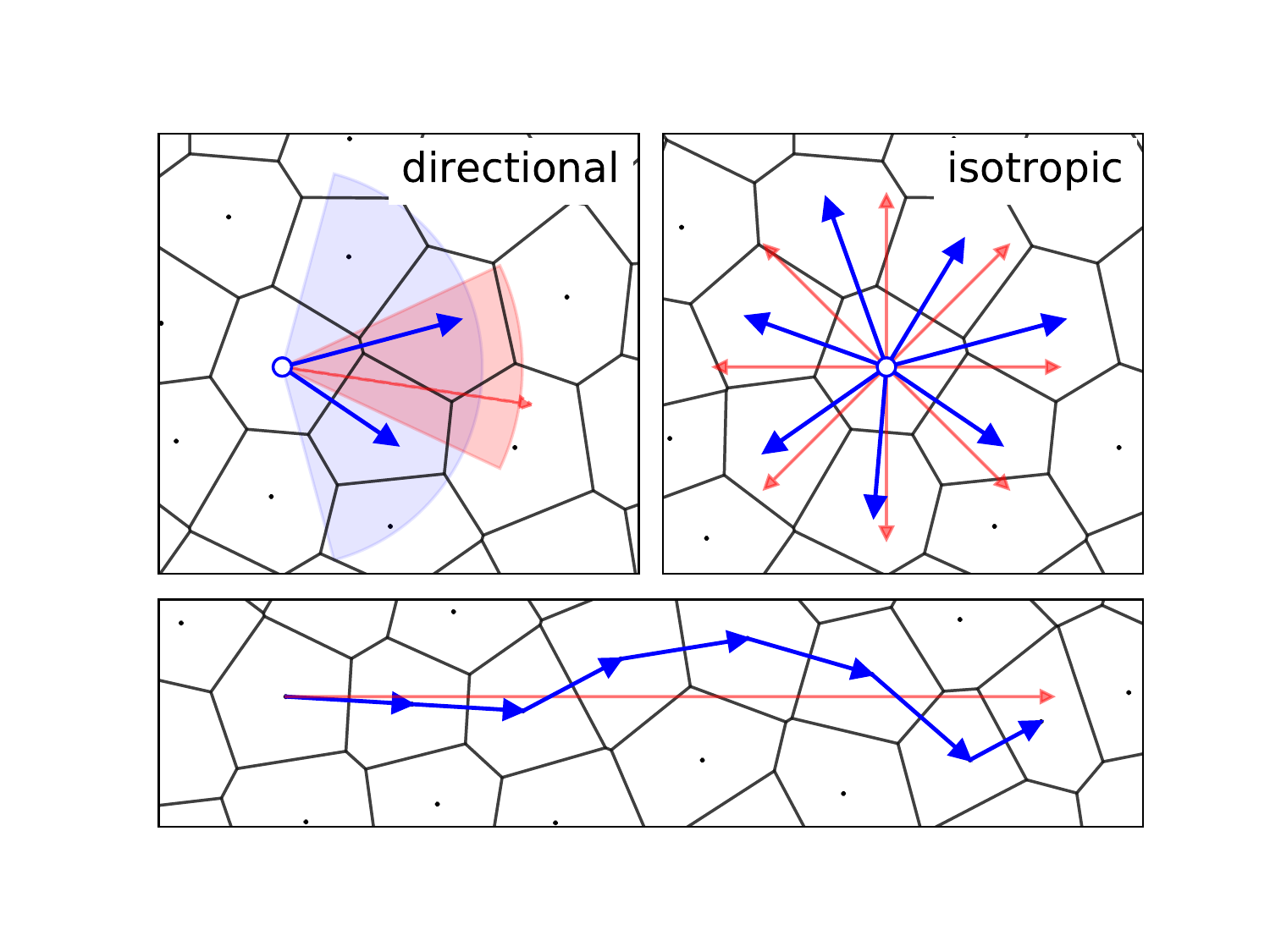}
    \caption{The two types of the photon transport used in SPRAI, directional and isotropic. For ease of visualization, the methods are shown here using a 2D Voronoi mesh, but they work in a similar fashion in 3D. Red arrows denote directional bin vectors and blue arrows show photon paths during the transport. The red cone depicts the solid angle $\Omega_{\mathrm{dir}}$ of the directional bin from which a random directional vector is chosen. The blue cone is the solid angle $\Omega_\mathrm{\theta}$ from which we choose most straightforward photon paths along the directional vector. The bottom panel shows a possible sequence of photon paths along the original direction.}
    \label{fig:photon_transports}
\end{figure}

The number of photons associated with each mesh cell, as well as the ionization rates and radiation fluxes are stored directly on the mesh. Each cell of the mesh has information about the number of photons that are passing through its volume, as well as their direction. For numerical convenience we define a set of $N_\mathrm{dir}$ solid angles (directional bins) equally distributed over the sphere using  simulated annealing. During the transfer, photons are moved from a directional bin of an initial cell to the corresponding directional bin of a destination cell. The photons are spatially moved, but their direction stays the same. We distinguish two types of photon transport, directional and isotropic. Both types are schematically described for a 2D Voronoi mesh in Figure \ref{fig:photon_transports}; the behaviour in 3D is similar, but harder to visualize.

\textbf{Directional transport} is used to transfer photons in a particular direction. Since the Voronoi mesh is not regular, it is not possible to move photons on straight lines defined by the directional vectors. Instead, we look for the most straightforward path along the particular directional bin (bottom panel). First we randomly choose a directional vector (red arrow) from the given solid angle $\Omega_{\mathrm{dir}}$ (red cone). This is done in order to prevent preferred paths on the grid and add some scatter to the photon beam. Then we look for three\footnote{For clarity, in Figure \ref{fig:photon_transports} we show only two neighbour paths.} neighbour cells for which the angles between their path vectors (blue arrows) and the directional vector are the smallest, and redistribute initial photons between them. By varying the solid angle $\Omega_\mathrm{\theta}$ (blue cone) we can exclude those paths which diverge too much from the original direction\footnote{This can happen if the number of directional bins $N_{\mathrm{dir}}$ is small,  as in that case $\Omega_{\mathrm{dir}}$ is large. In general, it is enough to exclude only backwards directions and set $\Omega_\mathrm{\theta}=2\pi$ sr.}.

\textbf{Isotropic transport} is similar to directional transport, but it is performed only for cells with sources. At the beginning of a RT step we calculate the number of new photons from the source emission rate and distribute them equally into each directional bin of a cell. In this case we compute random directional vectors (red arrows) for each directional bin (cones are not shown). Then we find all neighbourhood cells and compute their path vectors (blue arrows). In the last step we find for each directional vector the closest path vector and assign its photons to the corresponding directional bins of a neighbour cell.

\subsection{Radiation field}
\label{section:radiation_field}

In contrast with the original SimpleX code that updates chemistry after each photon step, our implementation of the SPRAI code first calculates a static ionization rate field and then passes it to an independent chemistry module. The module is similar to the one used in the Fervent code of \citet{Baczynski2015a}. We divide ionizing photons into two energy bins:
\begin{itemize}
	\item[] 13.6 eV < $\gamma_{13.6+}$ < 15.2 eV,
	\item[] 15.2 eV < $\gamma_{15.2+}$
\end{itemize}
Photons in the 13.6+ bin can ionize atomic hydrogen but not molecular hydrogen. Photons in the 15.2+ bin can ionized both H and H$_{2}$. We therefore have three possible reactions:
\begin{itemize}
	\item[] $\mathrm{H} \mbox{ }+ \gamma_{13.6+} \; \rightarrow \; \mathrm{H^{+}} + e^{-}$
	\item[] $\mathrm{H} \mbox{ } + \gamma_{15.2+} \; \rightarrow \; \mathrm{H^{+}} + e^{-}$
	\item[] $\mathrm{H_2} + \gamma_{15.2+} \; \rightarrow \; \mathrm{H^{+}_2} + e^{-}$.
\end{itemize}
We further assume \citep[as in][]{Baczynski2015a} that all of the H$_{2}^{+}$ ions formed by the third reaction are immediately destroyed by dissociative recombination:
\begin{itemize}
	\item[] $\mathrm{H_2^{+}} + e^{-} \; \rightarrow \; \mathrm{H + H}$.
\end{itemize}
We note that it is easy to extend the number of energy bins to treat e.g.\ helium photoionization, or the ionization of metal atoms in the present-day ISM. The method can also be extended to model the photodissociation of $\mathrm{H_2}$ by Lyman-Werner photons. However, the test problems in this paper are restricted to the simple ionization of hydrogen gas by photons with energies above 13.6 eV. Multifrequency tests will be included in subsequent papers.

As already discussed in the previous section, the chemistry of the gas is computed for any active mesh cells on each hydrodynamical sub-step, whereas the radiation field is updated only on full hydrodynamical steps $t_\mathrm{HD}$. To account for this, we could in principle introduce an additional restriction on the Arepo timestep to prevent the ionization state of the gas from changing too quickly. In practice, we have not found this to be necessary for the tests presented in this paper.

Photons in the system do not keep information about their previous path and therefore their attenuation in the gas has to be calculated locally for each cell. 
The number of photons $N^\gamma_\alpha$ from the frequency bin $\gamma$ that are attenuated by the species $\alpha$ is calculated as
\begin{equation}
 N^{\gamma}_\alpha = N^{\gamma}_\mathrm{in} \left( 1 - e^{-d\tau^{\gamma} } \right)  \frac{ \langle \sigma_{\alpha}^{\gamma} \rangle x_\alpha }{ \Sigma^\gamma },
\end{equation}
where $N^{\gamma}_\mathrm{in}$ is the number of photons coming in to the cell, $\langle \sigma_{\alpha}^{\gamma} \rangle$ is the mean absorption cross-section, $x_\alpha$ is the mass fraction of the species in the cell and the optical depth
\begin{equation}
d\tau^{\gamma} = dn \Sigma^\gamma
\end{equation}
 is summed over all species $\Sigma^\gamma \equiv \sum_\alpha \langle \sigma_{\alpha}^{\gamma} \rangle x_\alpha$. We calculate values for the mean cross-sections $\langle \sigma_\alpha^\gamma \rangle$ and  mean deposited energies $\langle E^\gamma \rangle$ (see below) in the same way as in \citet{Baczynski2015a}. However, in the tests in this paper, we use fixed values for these quantities, for simplicity. The column density of nucleons in the cell
\begin{equation}
dn = n_\mathrm{cell} \langle dr \rangle
\end{equation}
is calculated from the mean distance\footnote{We can assume spherical shapes of cells since the particle mesh in Arepo code is regularised. All mesh-generating points are automatically moved to the center-of-mass of their cells, so that the shape of cells is more regular and volumes of the neighbours are very similar.} $\langle dr \rangle$ to the neighbouring cell and the nucleon number density
\begin{equation}
n_\mathrm{cell} = \frac{\rho_\mathrm{cell}}{m_\mathrm{p}(1+4 \,Y_\mathrm{He})},
\end{equation}
where $\rho_\mathrm{cell}$ is the mass density of the cell, $m_\mathrm{p}$ is the proton mass, and $Y_\mathrm{He}$ is the fractional abundance of He in the gas. 

From the number of attenuated photons $N^\gamma_\alpha$ and the number of species $\alpha$ available for the reaction in the cell volume $V_\mathrm{cell}$ at the beginning of the RT 
\begin{equation}
\mathcal{N}_\alpha = n_\mathrm{cell} \, x_\alpha \, V_\mathrm{cell},
\end{equation}
we can calculate the corresponding reaction rate
\begin{equation}
 k_\alpha = \sum_\gamma k^\gamma_\alpha = \sum_\gamma \frac{N^{\gamma}_\alpha}{\mathcal{N}_\alpha} \frac{1}{t_\mathrm{HD}},
\end{equation}
where $t_\mathrm{HD}$ is the time of the previous full hydrodynamical step. Given the average deposited energy $\langle E^\gamma \rangle$ for the reaction, we can also calculate for each species a total heating rate
\begin{equation}
 \Gamma_\alpha =  \sum_\gamma k_\alpha^\gamma \langle E^\gamma \rangle .
\end{equation}
Values of $k_\alpha$ and $\Gamma_\alpha$ are stored for each cell and used by the chemistry module.

When dealing with photons in multiple direction bins entering a mesh cell, the fact that the equations above are linear in $N_{\rm in}^{\gamma}$ means that we can proceed in one of two ways. We  can compute individual values of $k_{\alpha}$ and $\Gamma_{\alpha}$ for each direction bin, and then simply sum them to get the final values for the mesh cell. Alternatively, we can sum the incident photons to get a single value of $N_{\rm in}^{\gamma}$ for the mesh cell and then use this to calculate $k_\alpha$ and $\Gamma_\alpha$ for the mesh cell, as above. The only complication in this case is the need to account for the fact that the photons lost in the cell, $N_{\alpha}^{\gamma}$, are spread over multiple directional bins. 

In our calculations we constrain the number of photons entering the reaction by the total number of species available for reactions\footnote{It is necessary to mention that the total number of available species for reaction should also include species created during the recombination in the cell. Since the recombination is treated later in the chemistry module we neglect it at this point. Nevertheless, for large $t_\mathrm{HD}$ this contribution might become important.}
\begin{equation}
N^\gamma_\alpha =
\begin{cases}
    N^\gamma_\alpha    & \mathrm{if} \;\; N^\gamma_\alpha < \mathcal{N}_\alpha \\
    \mathcal{N}_\alpha  & \mathrm{if} \;\; N^\gamma_\alpha \geqslant \mathcal{N}_\alpha, \\
\end{cases}
\end{equation}
where $\mathcal{N}_\alpha$ is the number of available species $\alpha$ in the cell at the current RT step. If the number of photons entering the cell $N^\gamma_\mathrm{in}$ is larger than the number of atoms or molecules available to ionize, the excess photons are transferred together with the unattenuated photons to the next cell.

The calculations described in this section are photon conserving. However, in order to enhance the computational performance of the code we stop calculations when the number of photons is too small. If the number of photons leaving the cell $N_\mathrm{out}^\gamma = N^\gamma_\mathrm{in} - \sum_\alpha N^\gamma_\alpha$ is smaller than some threshold value $N_\mathrm{th}$, then we do not transfer them further to the neighbouring cells, but instead remove them from the system. In our simulations we set the threshold value four orders of magnitude lower than the number of available species in the particular cell, so that $N_\mathrm{th} \ll \sum_\alpha \mathcal{N}_\alpha$ and the photon loss is minimized.

\subsection{Chemistry}
The chemical evolution of the gas is calculated separately for each cell after the end of the RT step, using as an input the photoionization rates calculated during the RT step (see Section~\ref{section:radiation_field} above). To model the chemical evolution, we use the primordial chemistry network that is available as part of the SGChem module in Arepo. This network was originally implemented in Arepo by \citet{hartwig15} and is based on earlier work by \citet{gj07}, \citet{ga08}, \citet{clark11} and \citet{glo15}. We also include several improvements to the treatment of He$^{+}$ recombination and H$^{-}$ photodetachment described in \citet{schauer17}. As previously noted, when we include H$_{2}$ photoionization in this network, we follow \citet{Baczynski2015a} and assume that in the conditions in which this process is important, all of the resulting H$_{2}^{+}$ ions will be almost immediately destroyed by dissociative recombination, yielding atomic hydrogen, so that the net effect is:
\begin{equation}
{\rm H_{2} + \gamma} \rightarrow {\rm H_{2} + e^{-}} \rightarrow {\rm H + H}.
\end{equation}
We solve for the thermal evolution of the gas due to radiative heating and cooling at the same time as the chemical evolution. We use the cooling function presented by \citet{ga08} and updated by \citet{glo15}. We also account for the heating due to the photoionization of H and H$_{2}$, which is computed as described in Section~\ref{section:radiation_field}.

We note that in principle, SPRAI is not restricted to modelling photoionization in metal-free gas and that it would be a relatively simple matter to couple the scheme to a present-day chemical model of the kind used e.g.\ in \citet{Baczynski2015a}. However, this lies outside the scope of the present paper.

\section{Tests and results}
\label{sec:TestsAndResults}

\begin{table*}
 \begin{tabular}{llcccccccccc}
    \multicolumn{2}{l}{Test} & $T_\mathrm{gas}$ [K] & $n_\mathrm{H}$ [$\mathrm{cm^{-3}}$] & $L_\mathrm{box}$ [pc] & $\gamma$ & $\dot{N}$ [ph/s]  & $t_\mathrm{rec}$ [yr] & $R_\mathrm{St}$ [pc] \\
    \hline
    1 & R-front     & 100  & 0.001 & $1.28\times10^4$ & 1.0001 & $10^{49}$                    & $1.22\times10^8$  & $6.79\times10^3$  \\
    2 & D-front     & 100  & 0.001 & $3.2\times10^4$  & 5/3         & $5 \times 10^{48}$                     & $1.22\times10^8$  & $5.39\times10^3$ \\
    3 & $r^{-2}$   & 100   & 100    & 13                       & 1.0001    & $10^{49}$                   & $1.22\times10^3$  & 3.15 \\
    4 & Blob test  & 1000 & 1       & 32                        & 5/3          & $1.62\times10^{48}$  & $1.22\times10^5$  & 36.95 \\
    \hline
 \end{tabular}
 \caption{Summary of the parameters used in different tests}
 \label{tab:simulation_summary}
\end{table*}

In this section, we present results from a set of standard tests to verify the accuracy of our implementation. The first test is a simple expansion of an HII region in a homogeneous environment. For this test, we disable photoionization heating and keep the gas temperature constant, allowing us to examine the behaviour of the ionization front (I-front) in the R-type regime. In the second test, we re-enable photoionization heating and allow the gas temperature to vary, enabling us to study the behaviour of the I-front in the D-type regime, where the propagation of the front is sensitive to the dynamical response of the gas. The setup of the third test is similar to Test 1, but instead of a homogeneous environment we select a radial density gradient proportional to $1/r^2$ for the gas (a setting similar to the Bonnor-Ebert sphere). In the last test we study the formation of a shadow behind a dense clump irradiated by two sources. Initial parameters and values for all of our test simulations are summarised in Table~\ref{tab:simulation_summary}. At the end of this section we discuss comparison of simulation times for different numbers of sources and resolutions.

In all of the tests, we use a directional base with $N_\mathrm{dir}=128$ bins, solid angle $\Omega_\Theta\mathrm{=2\pi}$~sr and the ionization rate is integrated from $N_{\rm rot}=5$ rotations\footnote{Rotations of the field are explained in the Appendix~\ref{sec:simplexRays}.} of the directional base. The mesh cells in our initial conditions are created using uniform Poisson sampling, and the mesh is relaxed prior to the run, so that all cells have approximately similar masses and shapes. For the tests below, we do not use any mesh refinement criteria, but in principle these could easily be combined with the basic algorithm.

\subsection{Test 1: R-type expansion of an HII region}

In the first test we look at the R-type expansion of a single HII region in a homogeneous gas environment with four different resolutions: $32^3$, $64^3$, $128^3$ an $256^3$ cells. (For brevity, we refer to these hereafter simply as resolutions of 32, 64, 128 and 256, respectively).

Expansion of an HII region is a well known problem. \citet{Stromgren1939a} showed that in ionization equilibrium, a point source of ionizing photons in an uniform density gas of pure hydrogen creates a spherical HII region with radius:
\begin{equation}
R_\mathrm{St} = \left[ \frac{3 \dot{N}_\gamma}{4\pi\alpha_\mathrm{B} (T) n_\mathrm{e}^2} \right]^{1/3},
\end{equation}
where $\alpha_\mathrm{B} (T)$ is the case B recombination coefficient of hydrogen, and $n_{\rm e}$ is the electron number density. If the gas inside the HII region is almost entirely ionized, then $n_{\rm e} \simeq n_{\rm H}$, the H atom number density of the undisturbed gas. Provided that the density of the ionized gas remains approximately constant, the time evolution of the ionization front (I-front) radius can be parametrised as
\begin{equation}
R_\mathrm{I}(t) = R_\mathrm{St} \left( 1 - e^{ -t/t_\mathrm{rec} } \right),
\label{eq:ionization_front}
\end{equation}
where the recombination time is calculated as
\begin{equation}
t_\mathrm{rec} = [ \alpha_\mathrm{B} (T) n_\mathrm{H} ]^{-1}.
\end{equation}

In order to directly compare our implementation with the Fervent code we use the same initial conditions as in the R-type test of \citet{Baczynski2015a}. Our simulation runs in a 3D box of size $L_\mathrm{box}=12.8$ kpc with periodic boundary conditions. A single idealised source with photon emission rate $\dot{N}_\gamma=1\times10^{49}$ photons per second is located at the centre of the box. The box is filled with a homogeneous gas with number density $n_\mathrm{H}=10^{-3} \; \mathrm{cm^{-3}}$. We assume that the initial gas in the box consists only of neutral hydrogen ($x_{\mathrm{H^{+}}}=0$) and has initial temperature $T_\mathrm{i}=100$ K. Since in this test we are not interested in the density response of the gas, we assume a quasi-isothermal equation of state where $\gamma=1.0001$, to keep the pressure constant across the ionization front. Besides that, we also fix the recombination rate coefficient to $\alpha_\mathrm{B}=2.59\times10^{-13}$ cm$^3$ s$^{-1}$ and the photoionization cross-section of the hydrogen to $\sigma_\mathrm{H}=6.3\times10^{-18}$ cm$^2$. These values lead to a Stro\"{o}mgren sphere with radius $R_\mathrm{st}\approx6.79$ kpc and the recombination time $t_\mathrm{rec}\approx122.34$ Myr.

\begin{figure}
  \includegraphics[width=\columnwidth,trim={0cm 0.2cm 0.7cm 0.8cm},clip]{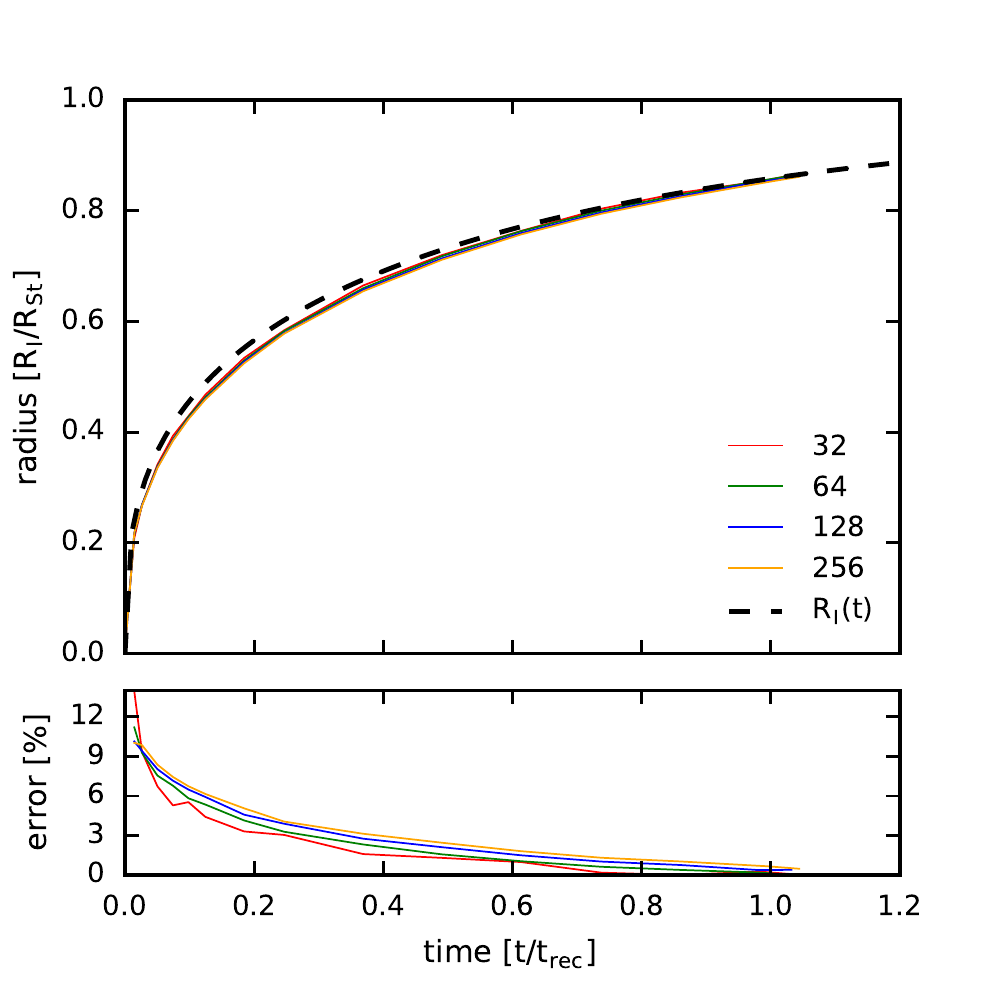}
    \caption{Results of test 1: time evolution of the radius of the HII region, $R_\mathrm{I}$ (top) and its deviation from the analytical solution (bottom). Coloured lines represent results of our simulations with different resolutions: 32, 64, 128 and 256. The black dashed line shows the values given by Equation \ref{eq:ionization_front}. }
    \label{fig:RfrontRad}
\end{figure}

The time dependent evolution of the I-front for all four resolutions (solid lines) as well the analytic solution (dashed line) for the I-front radius $R_\mathrm{I}$ are shown in Figure \ref{fig:RfrontRad}. In the bottom panel, we plot the relative error between the simulated and analytical radius. From the figure we can see that the I-front in the simulations closely follows the analytical solution. Significant errors are only visible in the initial stage but with increasing time this difference decreases. This is presumably due to the relatively coarse time resolution of the radiative transfer, which is restricted by the time interval of the highest hydrodynamical step. Nevertheless, our results match very well with the results of \citet{Baczynski2015a}.

\begin{figure*}
  \includegraphics[width=\textwidth,trim={2cm 0cm 1.4cm 0.7cm},clip]{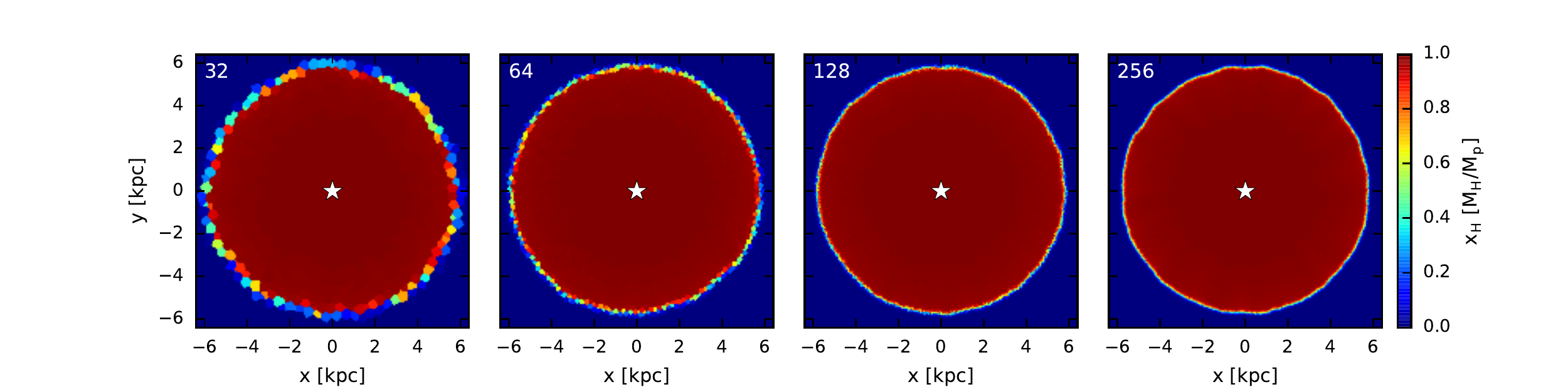}
    \caption{Results of test 1: mass fraction of the ionised gas in the z=0 plane. Results are for all simulated resolutions 32, 64, 128 and 256 at time t=75 Myr. The white star marks the position of the ionizing source.}
    \label{fig:RfrontProp}
\end{figure*}

In Figure \ref{fig:RfrontProp} we plot the mass fraction of ionised hydrogen in the plane $z=0$ for each resolution. In the lowest resolution case we can clearly distinguish the cell structure of the Voronoi mesh. This is especially visible at the edge of the ionization front. With increasing resolution, these features disappear and the I-front itself gets thinner. 

\subsection{Test 2: D-type expansion of an HII region}

In the previous test we simulated the quasi-isothermal case of the HII region expansion. In this test we allow ionising photons to heat the gas. Any change of the temperature increases the gas pressure around the source. Initially, the ionization front propagates so rapidly that the gas has no chance to respond hydrodynamically to this increase in pressure. However, once the speed of the I-front drops below 2$c_\mathrm{s}$, where $c_\mathrm{s}$ is the speed of sound in the ionized gas, the expansion of the gas causes a shock to separate from the I-front and proceed it into the surrounding neutral gas. Once this occurs, we describe the ionization front as a D-type front.

\begin{figure}
  \includegraphics[width=\columnwidth,trim={0cm 0.3cm 0.7cm 0.7cm},clip]{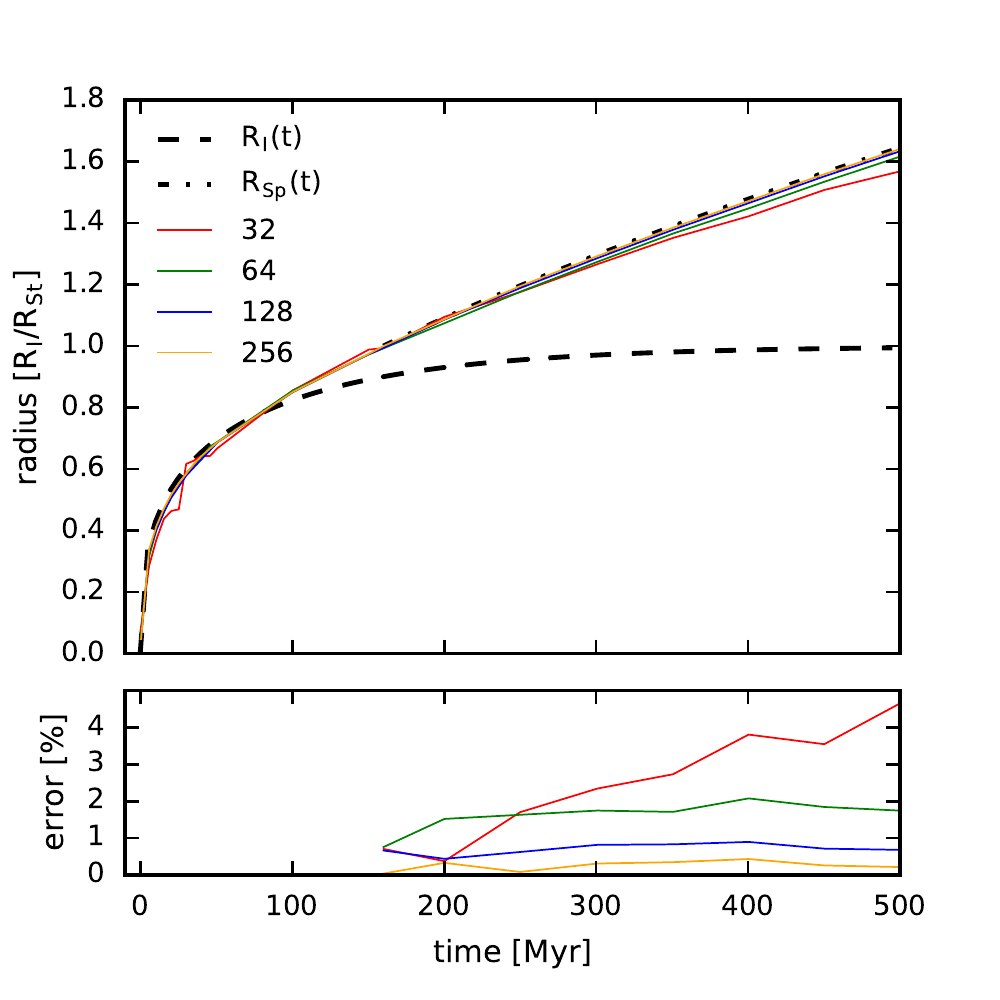}
    \caption{Results of test 2: Time evolution of the D-type I-front for resolutions 32, 64, 128 and 256. The black dashed line shows the values given by Equation \ref{eq:ionization_front}. The black dashed-dotted line is the Spitzer radius, computed assuming an average HII region temperature $T_\mathrm{avg}=1.1\times10^{4}$ K. The Spitzer solution is calculated from the time $t\approx160$ Myr when the radius of the I-front first reaches the Str\"{o}mgren radius. In the bottom panel, we plot the fractional amount by which each simulation differs from the analytical Spitzer solution.}
    \label{fig:DfrontRad}
\end{figure}

Due to the dynamical response of the matter, the position of the I-front is no longer given by Eq. \ref{eq:ionization_front}. Instead it enters as \citep{Spitzer1978}
\begin{equation}
\label{eq:spitzerEquation}
R_\mathrm{Sp}(t) = R_\mathrm{St} \left( 1 + \frac{7}{4} \frac{c_\mathrm{s} t}{R_\mathrm{St}} \right)^{4/7},
\end{equation}
where the speed of the sound $c_\mathrm{s}$ can be calculated from the average temperature $T_\mathrm{avg}$ in the ionised gas as
\begin{equation}
c_\mathrm{s} = \sqrt{ \frac{\gamma k_\mathrm{B} T_\mathrm{avg}}{m_\mathrm{H}} },
\end{equation}
where $k_\mathrm{B}$ is the Boltzmann constant and in case of a pure hydrogen gas with atomic weight $m_\mathrm{H}$ we use $\gamma=5/3$.

\begin{figure*}
  \includegraphics[width=\linewidth,trim={0.9cm 2.0cm 1.7cm 2.5cm},clip]{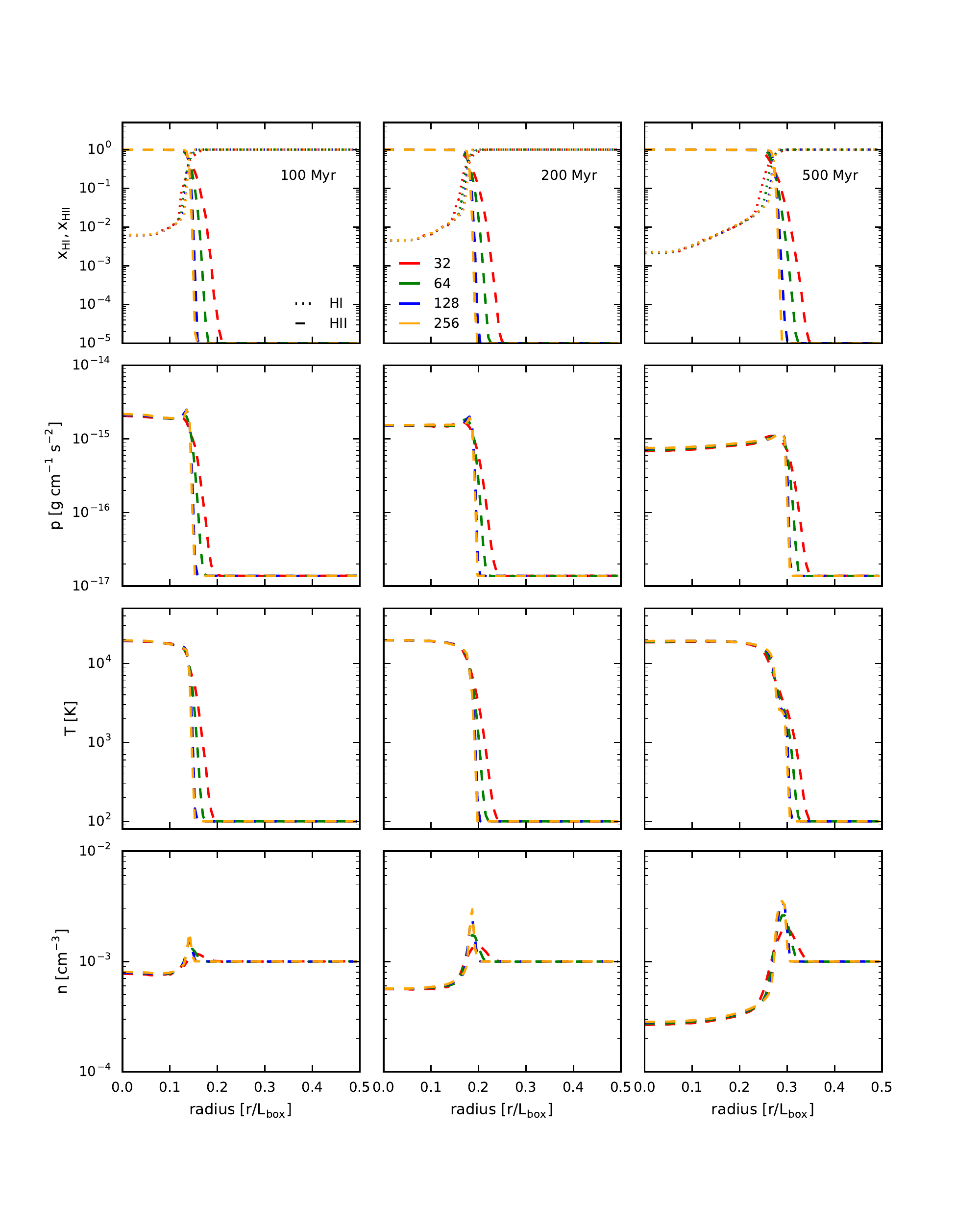}
    \caption{Results of test 2: various properties of the gas at times 100, 200 and 500 Myr as functions of distance from the star. In the first row we plot mass fraction of neutral and ionised hydrogen. The second, third and fourth rows show pressure, temperature and the proton density of the gas, respectively.}
    \label{fig:DfrontSphereProp}
\end{figure*}

\begin{figure*}
  \includegraphics[width=\linewidth,trim={1.8cm 2.4cm 0.9cm 2.6cm},clip]{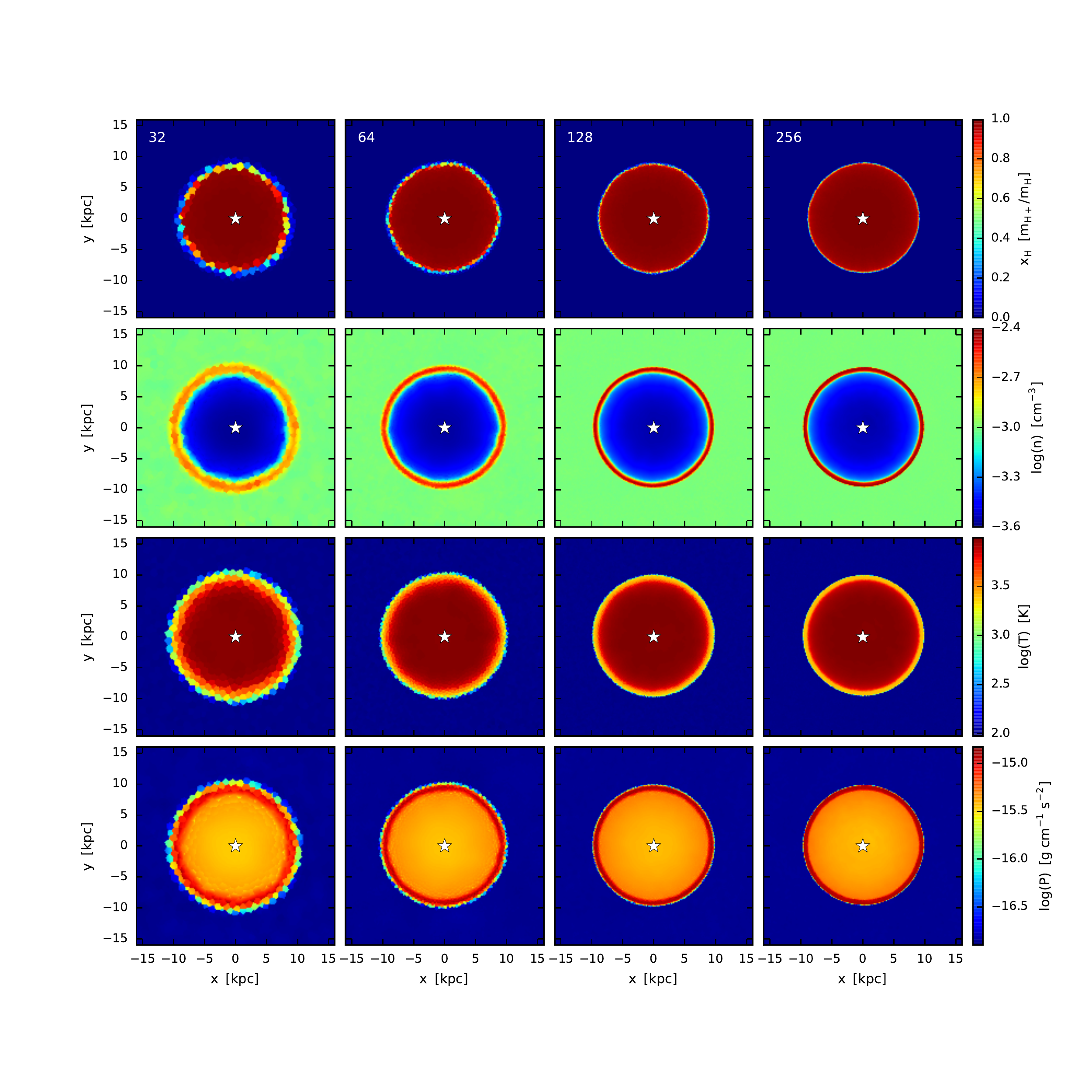}
    \caption{Results of test 2: slices through the simulated box in the plane $z=0$ at time $t = 500$~Myr. From the top to the bottom, the rows show the mass fraction of ionised gas, the temperature, the proton number density and the pressure of the gas, respectively. From left to right, the columns show results for different resolutions: 32, 64, 128 and 256, respectively.}
    \label{fig:DfrontProp}
\end{figure*}

Initial conditions for this test are similar to the D-front simulation (Test 5) of \citet{Iliev2009}. In our case we put an ideal ionization source $\dot{N}_\gamma=5\times10^{48}$ s$^{-1}$ into the centre of a box with $L_\mathrm{box}$ = 32 kpc \footnote{Note that \citet{Iliev2009} used $L_\mathrm{box}$ = 30 kpc.}. The box is filled with neutral hydrogen gas with initial density $n_\mathrm{H}=10^{-3}$ cm$^{-3}$ and temperature $T$ = 100 K. Each ionization event deposits an energy of 2 eV to the corresponding gas cell. The values of $\alpha_\mathrm{B}$ and $\sigma_\mathrm{H}$ are fixed to the same values as in the Test 1. The corresponding Str\"{o}mgren radius and recombination time for these values are $R_{\rm St}$ = 5.4 kpc and $t_{\rm rec}$ = 122.4 Myr, respectively. In order to see the later stages of the I-front evolution we run the simulation for $4 \times t_{\rm rec}\approx 500$ Myr.

The time evolution of the I-front for all four resolutions is shown on Figure \ref{fig:DfrontRad}. The dashed black line indicates the Str\"{o}mgren radius and the dashed-dotted line is the Spitzer solution. The coloured lines are results for each resolution, where the I-front is defined as a radial distance from the source where the ionization fraction of atomic hydrogen drops below 50\%. We find that the simulated data are in a good correspondence with the analytical solutions and results of \citet{Iliev2009}. Large errors for the run with resolution of 32 are caused by the relatively large size of the gas cells in comparison to the physical size of the shock front.

The next set of plots in Figure \ref{fig:DfrontSphereProp} show several gas properties in the box as a function of radial distance from the source at three different times: 100, 200 and 500 Myr. In the first column we plot mass fractions of neutral and ionised hydrogen. In the second, third and fourth rows we plot pressure, temperature and density, respectively. These results are also in good agreement  with the results of \cite{Iliev2009}. From all the plots we can see that the thickness of the shock decreases with increasing resolution. Grid cells for the two lowest resolutions, 32 and 64, are bigger than the physical size of the shock and therefore it is not well resolved. However, the size of the shock front settles on a constant value for the two highest resolutions. 

\subsection{Test 3: HII region expansion in a density gradient}

The previous two tests assumed that the radiation is transferred in an environment with homogenous density. This is, however, not the case for the astrophysical processes that we want to study. Stars are usually formed in the centre of a collapsing gas cloud with a strong radial density gradient. Therefore, in this test we focus on the idealised case used also in \citet{Baczynski2015a}. 

We place a source of radiation at the centre of a sphere with the density profile
\begin{equation}
 n(r)=
 \begin{cases}
    n_\mathrm{c} & \text{ if } r \leq r_\mathrm{c} \\
    n_\mathrm{c} (r/r_\mathrm{c})^{-2} & \text{ if } r > r_\mathrm{c},
 \end{cases}
\end{equation}
and let it evolve. We pick a central density of $n_c$ = 100 $\mathrm{cm^{-3}}$ and a core radius of $r_c = 1.97$~pc. For this choice of density profile it is possible to calculate an analytical solution \citep{Franco1990,Whalen2006,Mellema2006} for the radius of the I-front as
\begin{equation}
 R_\mathrm{grad}(t) = r_\mathrm{c} (1+ 2 t n_\mathrm{c} \alpha_\mathrm{B})^{1/2}.
\end{equation}
We take $t = 0$ to be the time when the ionization front reaches the edge of the dense core, i.e.\ $R_\mathrm{grad}(0)=r_\mathrm{c}$.

In this test, we again set $\gamma$=1.0001 to prevent the density response due to gas heating. The ionization source produces $\dot{N}_\gamma=1\times10^{49}$ photons per second. The recombination rate is fixed to a value of $\alpha_\mathrm{B}=2.59\times10^{-13} \mathrm{cm^3}$ s$^{-1}$ and the photoionization cross-section of the helium to $\sigma_\mathrm{H}=6.3\times10^{-18}$ cm$^2$.

\begin{figure}
  \includegraphics[width=\columnwidth,trim={0.2cm 0.1cm 0.9cm 0.8cm},clip]{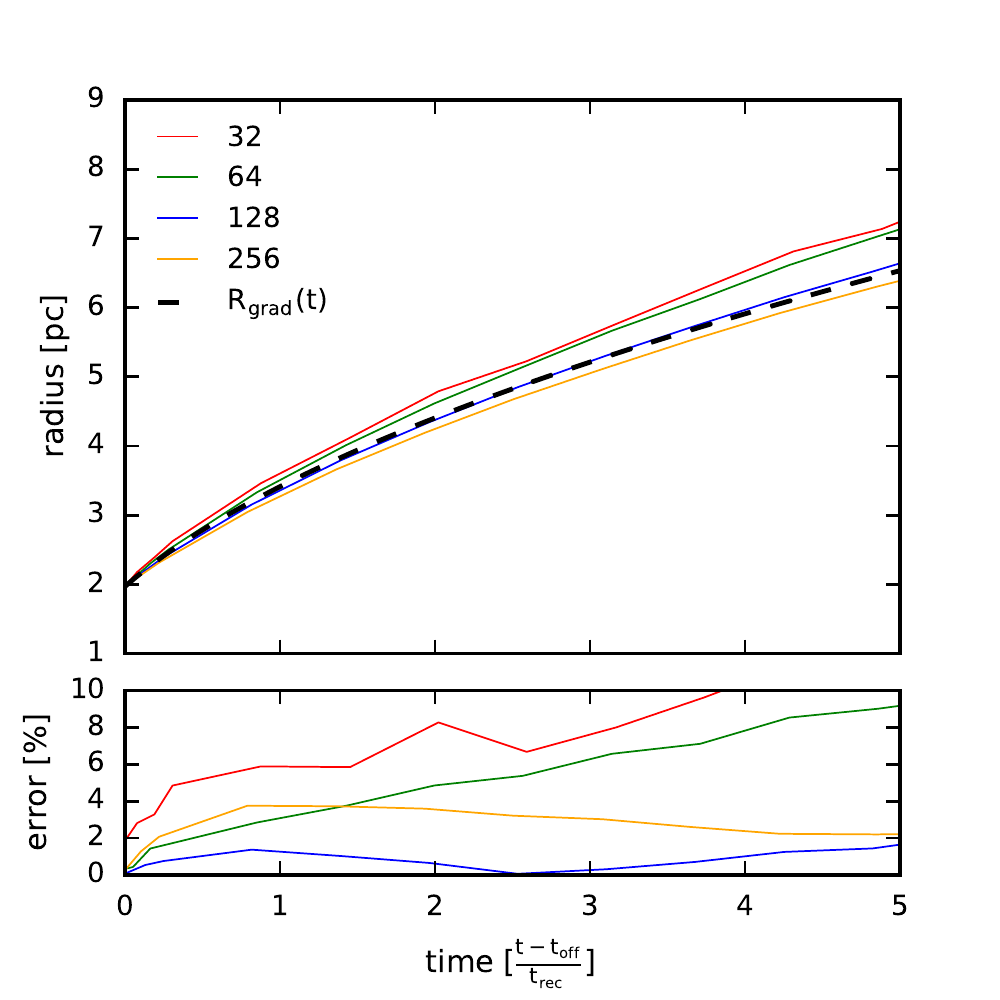}
    \caption{Results of test 3: time evolution of the R-type I-front for resolutions 32, 64, 128 and 256 in gas with a radially decreasing density profile. The black dashed line is the analytical solution calculated for this test.  The coloured lines are results of the simulations starting from $\mathrm{t_{off}}$ when the I-front reaches the core radius $r_{\rm c}$. In the bottom panel we plot relative errors for each simulation compared to the analytical values.}
    \label{fig:GRfrontRad}
\end{figure}

Results of this test are shown in Figure \ref{fig:GRfrontRad}. The duration of the simulation is approximately five times the recombination time ($5 \times t_\mathrm{rec} \approx 6.73$ kyr) in the dense core. 

We define the radius of the I-front as the distance from the source where the fraction of ionised hydrogen drops below 90\% of the total mass of the gas in the corresponding spherical shell. From the plots we can see that the two lowest resolutions, 32 and 64, have the largest errors. The grid cells in these two simulations are not small enough to properly resolve density gradients at large radii and therefore the shapes of their I-fronts are not entirely spherically symmetric. Instead, they depend on the mass inhomogeneities caused by the grid. Simulations with higher resolution give proper radially symmetric I-fronts that evolve according to expected values.

\subsection{Test 4: Photo-evaporation of a dense clump by two sources}

\begin{figure*}
  \includegraphics[width=\linewidth,trim={1.9cm 2.3cm 0.9cm 2.6cm},clip]{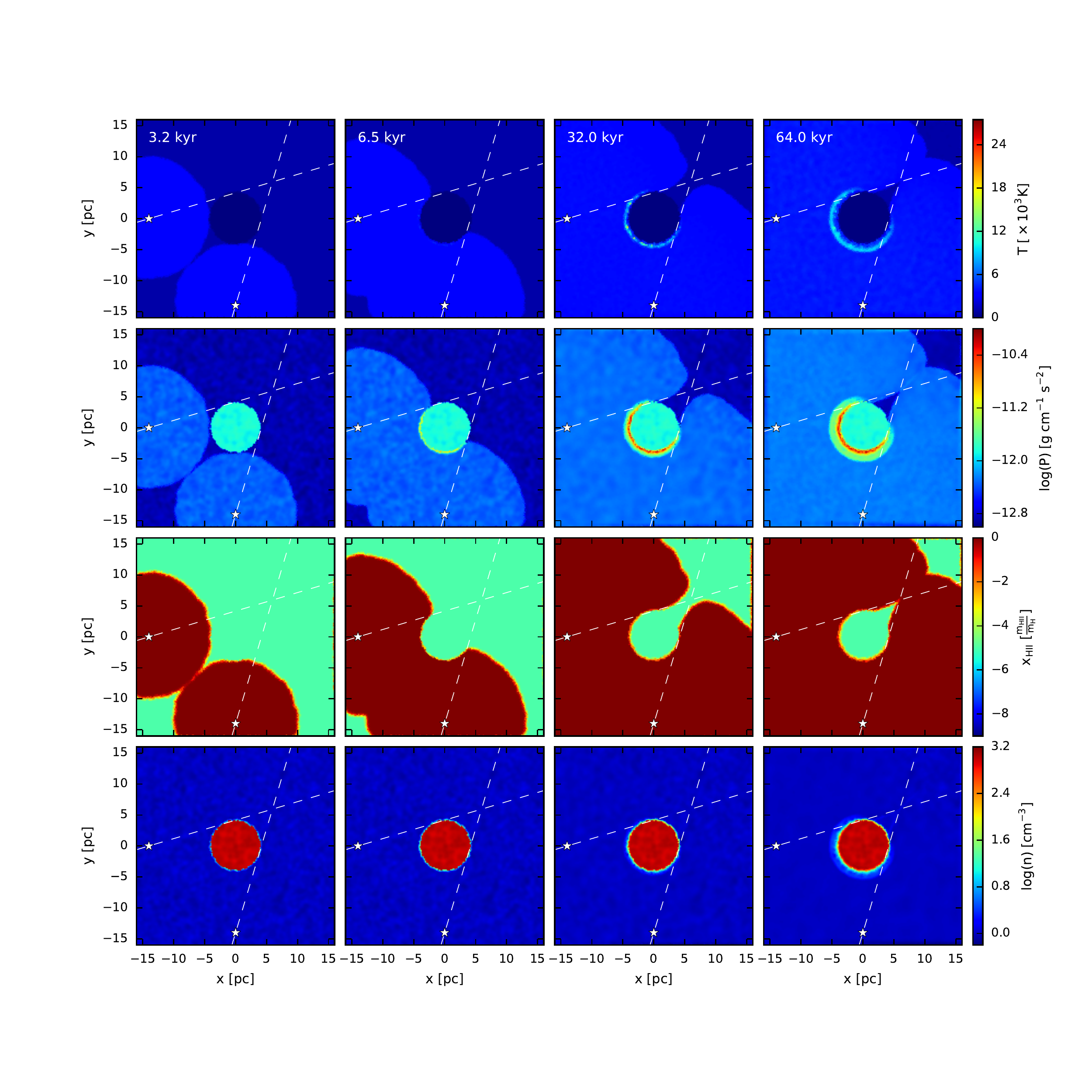}
    \caption{Results of test 4: this figure shows the formation of the shadow behind a dense clump placed in the centre of the simulation box irradiated by two sources marked by stars. From left to right, the columns show $z=0$ slices through the box at output times 3.2, 6.5, 32 and 64 kyr, respectively. From the top to the bottom, the rows show gas temperature, pressure, the mass fraction of the ionised gas and the total density of the gas, respectively. Dashed white lines are added to show the boundaries of the expected shadow behind the clump.}
    \label{fig:BlobProp}
\end{figure*}

In this test we show how well our method resolves the formation of shadows behind dense objects. Our initial settings are similar to \citet{Baczynski2015a} (see their Section 3.6). The difference is that here we use only a hydrogen ionising frequency bin. A box of size $L_\mathrm{box}=32$ pc is filled with neutral hydrogen gas with density $n=1 \, {\rm cm}^{-3}$ and temperature 1000 K. In the middle of the box we placed a dense clump of gas with $n=1000 \, {\rm cm}^{-3}$, radius $r_\mathrm{clump}=4$ pc and temperature 10 K. The clump is irradiated by two sources located at positions $p_1$($x$,$y$,$z$) = (-14, 0, 0) pc and  $p_2$($x$,$y$,$z$) = (0, -14, 0) pc. Each of these sources emits $\dot{N}_\gamma=1.61\times 10^{48}$ photons per second. The photoionization cross-section of hydrogen is fixed to $\sigma_\mathrm{H}=5.38\times10^{-18}$ cm$^2$ and each photoionization heats the gas by 0.72 eV. The recombination rate is as usual fixed to value $\alpha_\mathrm{B}=2.59\times10^{-13}$ cm$^3$ s$^{-1}$. The simulation runs until both ionization fronts pass around the dense clump (64 kyr) and form a shadow of non-ionised gas behind it.

In Figure \ref{fig:BlobProp} we present the results of the simulation with resolution 128\footnote{We show results for resolution 128, because our simulation with resolution 256 did not finish in a reasonable time. For highly ionised environments, as in this test, communication between processors appears to be slow. We plan to address this problem in future work.} at four different times: in the first column, the I-fronts are about to intersect. In the second column, the two regions of ionised gas are already joined and start ionising the dense clump. In the third column the clump is being heated and we see the formation of a shadow behind it. In the last column, the shadow is clearly visible. Its contours are not aligned with the white dashed lines because photons of our rays can experience small side-shifts that ionise the cells also within the shadow region. This artefact of the SimpleX method can be reduced by increasing the resolution. 

The rows in Figure \ref{fig:BlobProp} correspond to different properties. The top row shows temperature, the middle row pressure and the two bottom rows illustrate the mass fraction of ionised hydrogen and its density. White dashed lines are added in order to indicate the edges of the shadow region behind the dense clump. Both from the pressure and temperature plot we can see that the part of the clump that faces the two sources is being heated and slowly evaporated, whereas parts in the shadow stay unaffected by the radiation.

\subsection{Performance scaling with number of sources}

\begin{figure}
 \begin{tabular}{llcccc}
 	\hline
    	\multicolumn{2}{l}{Resolution}          & 32 & 64 & 128 & 256 \\ 
	\multicolumn{2}{l}{Number of procesors}  & 16     & 32       & 160     & 320 \\ \hline
	\multirow{3}{*}{ \begin{minipage}{1.3cm} Run time [seconds] \end{minipage} } 
						     & 1 source     & 2311 & 10265 & 36433 & 268938 \\
	                                               & 10 sources  & 2525 & 13042 & 51925 & 411294 \\
	                                               & 100 sources & 2741 & 14492 & 58529 & 462979 \\ \hline
\end{tabular}
  \includegraphics[width=\columnwidth,trim={0.4cm 0.4cm 0.4cm 0.3cm},clip]{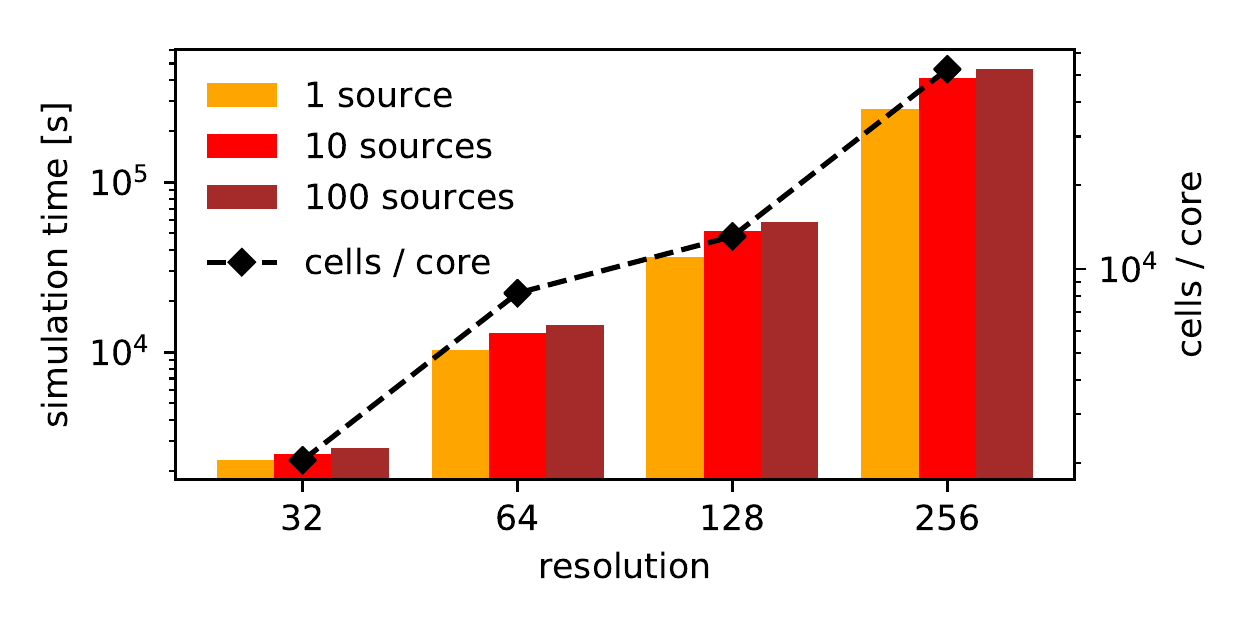}
    \caption{Dependance of the simulation time (in seconds) on the number of sources and the resolution. For simulations with resolutions 32, 64, 128 and 256, we used 16, 32, 160 and 320 cores, respectively. Times are calculated from time balance reports produced by the Arepo code.}
    \label{fig:SimTimes}
\end{figure}

A very common problem of ray-tracing radiative transfer simulations is that the required computation time increases linearly with the number of sources. However, this is not the case for the SimpleX method. Every cell in the grid can be considered as a source that radiates photons only to its neighbours. Therefore the main factor that influences the time required for a simulation is given by the resolution of the Voronoi grid and the maximum number of steps in the radiative transfer phase, which in turn is determined by the maximum number of cells photons can travel within the box. 

In order to show this advantage we performed a set of simulations of R-type I-fronts where we used 1, 10 and 100 sources, each with ionization rate $10^{49}$, $10^{48}$ and $10^{47}$  photons per second, respectively, so that in all three simulations the total production rate of ionising photons was the same. The setup of the simulation box is identical to Test 1. In simulations with 10 and 100 sources, we placed sources evenly on a spherical shell of radius $r\approx$ 700 pc around the original position of the single source. 

The total running times of simulations in seconds are summarised in Figure~\ref{fig:SimTimes}. From the results we can observe an increase of the time for larger source numbers. This time increase is, however, mainly related to the part of the code that searches gas cells near source locations and injects new photons into them, and the functional increase is small compared to the increase in the number of sources.

It is important to note that just as with other ray tracing algorithms, in SimpleX the computation time increases in regions with low radiation opacity. The time increase is caused by copying unattenuated photons from cell to cell. For example in Test 4, one can observe an sudden increase of the computational time when the two Str\"{o}mgren spheres meet each other, since at this point photons from the first source have to travel through the ionised region of the second source until they reach the ionization front on the other side, and vice versa.

\section{Conclusions}
In this paper, we introduced a new implementation of the SimpleX radiation transfer algorithm on the moving mesh of the hydrodynamical code Arepo. We refer to our combination of SimpleX and Arepo as SPRAI: Simplex Photon Radiation in the Arepo Implementation. We show the results when SPRAI is applied to several simple test problems that have analytical solutions: the expansion of an HII region in homogeneous gas in both the R-type and D-type regimes, and the expansion of an HII region in gas with a radially decreasing density profile.
All our tests produce results that are in good agreement with the analytical solutions and with the results of the ray-tracing code Fervent, which uses the same chemistry module.
In our last test, we irradiated a dense blob of gas with two identical sources and observed the formation of shadow behind it. This simulation is also in a good agreement with the results presented in \citet{Baczynski2015a}.
 
One of the main problems of this implementation is variation of the width of photon rays with different resolutions. This behaviour was already observed and described in the original SimpleX papers. In Appendix \ref{sec:simplexRays}, we show that for higher resolution simulations this has an important impact on the homogeneity of the calculated ionization rates. Our solution to this problem is to construct the ionization field from randomly rotated partial fields. This method  proved to be satisfactory for our testing problems, although it introduces additional demands on the computational time.

In the future we plan to optimise the performance of the code and perform some multifrequency tests that we did not include in this first paper. We would also like to address the problem with the photon ray spreading by introducing a continuous description of the radiation field using spherical harmonics. Nevertheless, our main goal is to use this code for real astronomical problems, including the study of the escape fraction of ionising photons from primordial mini-halos as well as simulations of the formation and fragmentation of accretion disks around Pop.\ III stars in the early universe.

\section*{Acknowledgements}
We thank Prof.\ Dr.\ Volker Springel for giving us access to the Arepo code and for his help. We also thank Paul Clark and Oliver Lomax for useful discussions on the behaviour of the SimpleX algorithm. We acknowledge support by the state of Baden-Württemberg through bwHPC and the German Research Foundation (DFG) through grant INST 35/1134-1 FUGG, as well as from the European Research Council under the European Community's Seventh Framework Programme (FP7/2007-2013) via the ERC Advanced Grant `STARLIGHT: Formation of the First Stars' (project number 339177). SCOG and RSK also acknowledge support from the DFG via SFB 881 ``The Milky Way System'' (sub-projects B1, B2, B8) and SPP 1573 ``Physics of the Interstellar Medium'' (grant number GL 668/2-1, KL 1358/18.1, KL 1358/19.2).



\bibliographystyle{mnras}
\bibliography{sprai} 




\appendix

\section{Behaviour of photon rays on a Voronoi grid}
\label{sec:simplexRays}

\begin{figure*}
  \includegraphics[width=\linewidth,trim={1.8cm 0cm 1.1cm 0.8cm},clip]{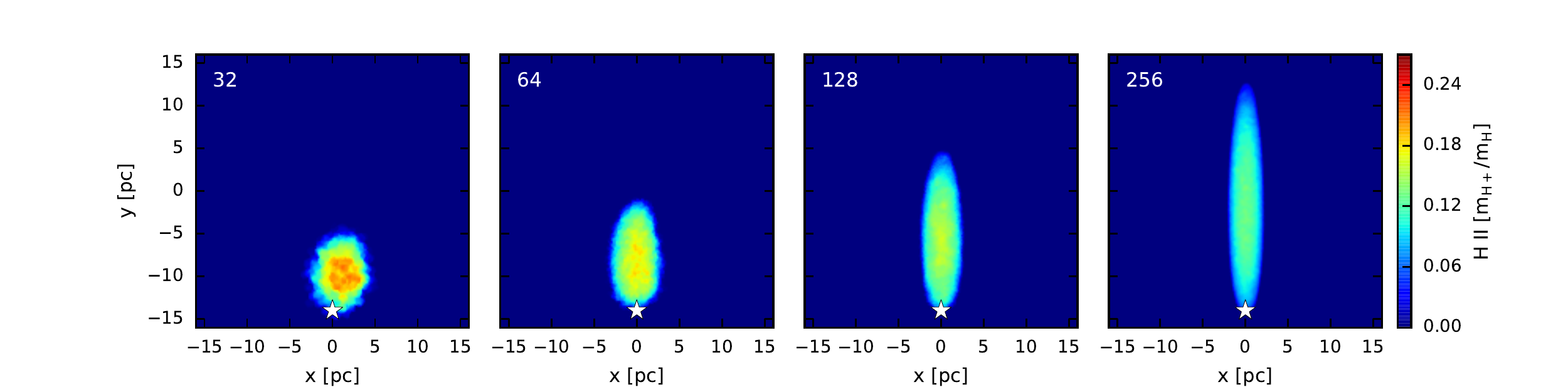}
    \caption{Mass-weighted projection of ionised hydrogen gas in the box produced by a single photon ray. Positions of sources are marked with stars and a single directional bin is orientated along the $y$-axis. Plotted rays are in an early stage of their growth. At later times all rays reach the upper border of the box, although their widths remain different.}
    \label{fig:RayIon}
\end{figure*}

\begin{figure*}
  \includegraphics[width=\linewidth,trim={0cm 0cm 0cm 0cm},clip]{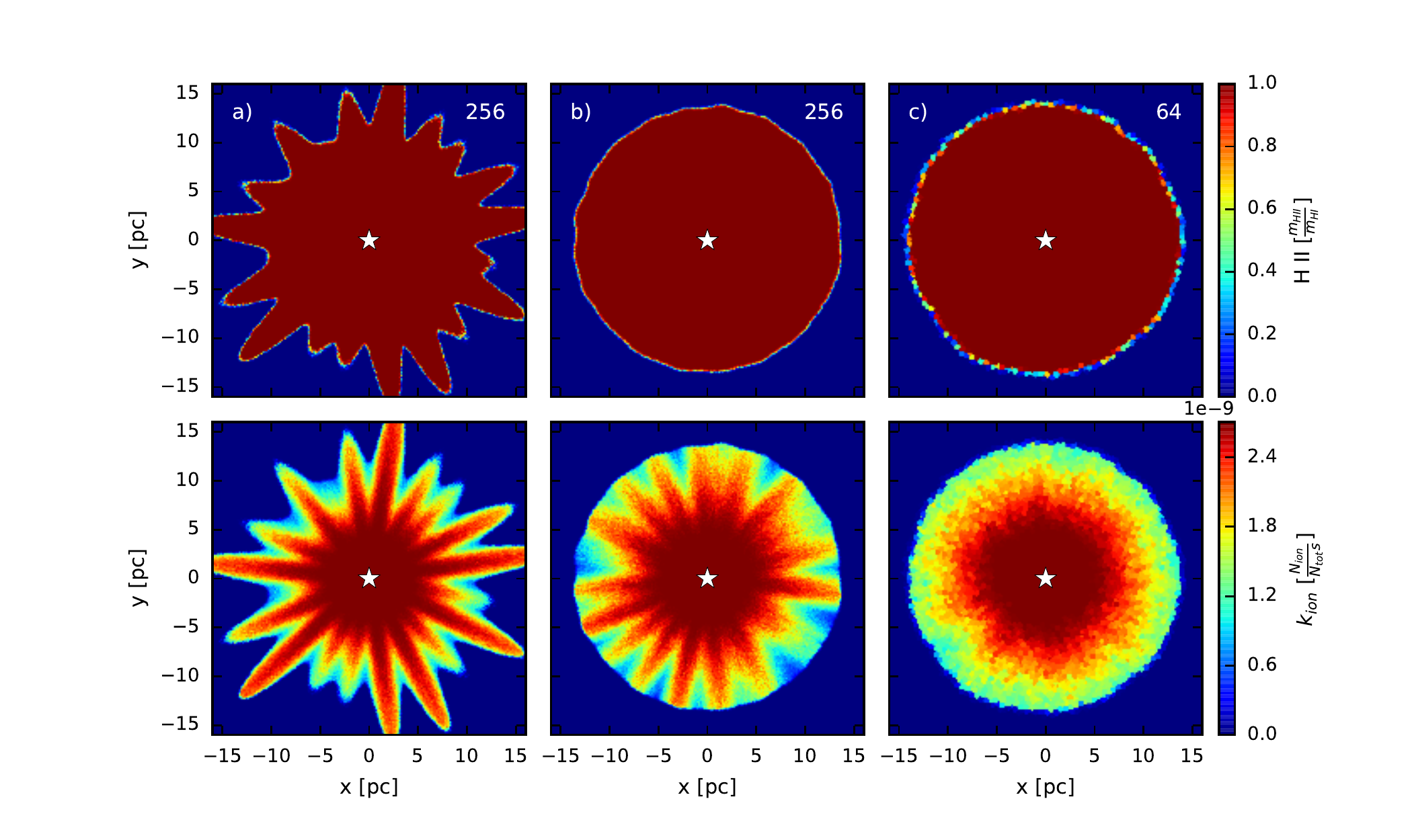}
    \caption{Effect of variable ray widths on the shape of a Str\"{o}mgren sphere. The top panels show the mass fraction of ionised gas and the bottom panels the corresponding ionization rate at time t=2.4 kyr for resolutions 256 and 64. In simulation a) a set of directional bins is fixed for all RT runs, whereas in simulations b) and c) the directional base is rotated five times per radiational transfer. The plots show slices through the box in the plane $z=0$.}
    \label{fig:CompareRfronts}
\end{figure*}

As mentioned in Section \ref{sec:TestsAndResults}, properties of the photon rays generated by SimpleX vary for different resolutions. In Figure~\ref{fig:RayIon}, we show results of a single ray ionization. In this setup we placed an ideal source similar to that in Test 1 at the bottom of a box filled with hydrogen gas, at coordinates (x,y,z) = (0, -14, 0). We enabled only one directional bin pointing in the direction of the $y$ axis and switched on the ionizing source. From the plot, we can see that the thickness and length of the ray depends significantly on the resolution. This is due to the fact that photons emitted from the sources are not moving on straight lines. Instead, they experience random side-shifts which depend on the size of the individual grid cells. Since the mean diameter of the cells for resolution 32 is eight times bigger than for resolution 256, photons rays are able to ionise gas further from the original directional axis. The length of the ray varies because low resolution cells contain more gas that can be ionised and therefore the same number of photons is used up closer to the source. 

This behaviour of rays at high resolutions results in an inhomogeneous distribution of photons far from the source and subsequent deformation of the I-front. In Figure \ref{fig:CompareRfronts}a, we plot a $z=0$ slice through the box of an R-type I-front simulation similar to Test 1 (but with a higher gas density and much smaller box) at time t=2.4 kyr and resolution 256. The top panel shows the ionised hydrogen fraction and the bottom panel the corresponding ionization rate. As we can see, although the I-front should be circular, given the symmetry of the problem, in  practice it  develops pointed features orientated parallel to the directional bins within the plane. At lower resolution, these feature are not present, because photon rays are able to overlap with each other (see Figure~\ref{fig:CompareRfronts}c). However, with increasing resolution, the radius of the overlap gets closer to the source and these features become visible. 

There are several ways to deal with this problem. One solution is to increase number of directional bins. The higher the number, the smaller the angle between two neighbouring photon rays, meaning that rays continue to overlap out to greater distances from the source. Nevertheless, it is important to note that using larger numbers of directional bins implies the
use of more memory and the need for more calculations per cell.

Another solution is to construct the total ionsisation field from a set of randomly rotated partial fields. In practice, this means that we divide the total number of photons at the beginning of the RT by $N_{\rm rot}$, and for each part perform separate radiation transfer. Summing the resulting $N_{\rm rot}$ fields gives the final field that is provided to the chemistry module. Results of a simulation using this method are plotted in Figure \ref{fig:CompareRfronts}b. Note that the ionization rate is still not completely homogeneous, but since the whole field is rotated $N_{\rm rot}$ times during each RT run, the time average of the resulting ionization rate is homogeneous and I-front features are effectively suppressed.  Nevertheless, this method requires more RT steps and therefore increases the required computational time. 


\bsp	
\label{lastpage}
\end{document}